\documentclass[12pt,a4paper]{article}

\usepackage{amsthm,amsmath,amsfonts,amssymb,url,graphicx,bm}

\usepackage{cite}
\usepackage{mathptmx,times}

\newcommand{\HH}{\mathcal{H}}

\newcommand{\CC}{\mathbb{C}}
\newcommand{\RR}{\mathbb{R}}
\newcommand{\ZZ}{\mathbb{Z}}

\newcommand{\GG}{\mathcal{G}}

\newcommand{\EE}{\mathcal{E}}
\newcommand{\VV}{\mathcal{V}}
\newcommand{\DD}{\mathcal{D}}
\newcommand{\PP}{\mathcal{P}}
\newcommand{\vc}[1]{\bm{#1}}
\newcommand{\vcm}[1]{\mathbf{#1}}

\newcommand{\be}{\mathrm{e}}
\newcommand{\bi}{\mathrm{i}}
\newcommand{\bd}{\mathrm{d}}

\newcommand{\dom}{\mathop{\mathrm{dom}}}
\newcommand{\diag}{\mathop{\mathrm{diag}}}

\newcommand{\spec}{\mathop{\mathrm{spec}}}

\newcommand{\Arccos}{\mathop{\mathrm{Arccos}}}

\newcommand{\indeg}{\mathop{\mathrm{indeg}}}
\newcommand{\outdeg}{\mathop{\mathrm{outdeg}}}

\newtheorem{thm}{Theorem}
\newtheorem{lemma}[thm]{Lemma}
\newtheorem{prop}[thm]{Proposition}
\newtheorem{corol}[thm]{Corollary}

\theoremstyle{definition}

\renewcommand{\emptyset}{\varnothing}

\begin{document}

{

\raggedright

\Large\bf

Localization effects in a periodic quantum graph
with magnetic field and spin-orbit interaction

\bigskip

\large\sc Konstantin Pankrashkin

\normalsize\rm

\bigskip

Institut f\"ur Mathematik, Humboldt-Universit\"at zu Berlin,
Rudower Chaussee 25, 12489 Berlin, Germany

\medskip

E-mail: \url{const@mathematik.hu-berlin.de}

\medskip

Homepage: \url{http://www.mathematik.hu-berlin.de/~const/}

}

\bigskip

\noindent {\bf Abstract.} A general technique for the study of magnetic
Rashba Hamiltonian in quantum graphs is presented.
We use this technique to show how manipulating the magnetic and spin parameters
can be used to create localized states in a certain
periodic graph ($T_3$ lattice).

\bigskip



\section{Introduction}

In this work, we discuss the creation of eigenvalues in periodic quantum graphs
by certain external interactions, namely, by magnetic field and spin-orbit
coupling.

The analysis of quantum graphs, i.e. of differential operators
on singular one-dimensional manifolds,  becomes one the central topics
in the mathematical physics during last decades, see~\cite{KS,KSM,kuch,Ku1,Ku2,mpp}.
This has many reasons; in particular, quantum graph Hamiltonians
appear in the de Gennes--Alexander theory of superconductivity~\cite{alx,castro,dGe}.
Some other fields of applications are described e.g. in~\cite{kuch}.

The spectral theory of compact quantum graphs has many common features
with the usual theory of differential operators, cf.~\cite{jvb2,nic,roth}.
Nevertheless, such an analogy is rather limited when considering
non-compact structures. Some particular
features of quantum graph models become obvious if one studies
periodic configurations. For example, for a large class of periodic
Schr\"odinger operators in Euclidian spaces the spectrum is known to be absolutely continuous~\cite{BirSus,asob}, while even simplest periodic quantum graphs
can have eigenvalues~\cite{Ku2,bgp}. Some other examples
may include the sensibility of periodic quantum graphs
to some arithmetic characteristics~\cite{exner-lkp}.

Recently, in the physics literature one discussed the so-called extreme localization
in the $T_3$ lattice (dice lattice)~\cite{p1,vid1}. From the mathematical point of view, it was shown that under certain magnetic fields the Hamiltonian of a quantum graph with the corresponding shape has no bands of continuous spectrum, and the spectrum
consists of infinitely degenerate eigenvalues.
This effect was observed also experimentally 
by transport measuring in superconducting and metallic wire networks~\cite{nap,naud1,naud2}.
(It is worth emphasizing that bound states in $T_3$ lattice appear without any external
interactions~\cite{suth}; the coexistence of the continuous and the point spectra
is implied by the rich internal symmetry of the lattice and of its dual, Kagom\'e~\cite{ae}.)
Various aspects of this  localization mechanism and its stability under disorder and external interactions are studied in a number of works~\cite{vid1,deloc,mc,vid2,r2}.
In particular, it is shown that additional interactions, like
the inter-particle interaction, destroy the extreme localization mechanism,
and continuous spectrum appears~\cite{deloc}.

In Ref.~\cite{r1} it was shown that in some periodic quantum graphs similar
localization phenomena can be induced not only by magnetic fields,
but also by spin-orbit interaction at certain values of the Rashba constant~\cite{br2,ra}.
Nevertheless, this analogy is limited, and the numerical analysis of~\cite{r2}
shows that the Rashba localization does not appear in the $T_3$ lattice.
In the present paper we consider the above situation
with both the spin-orbit and magnetic interactions.
We note that the quantum graph models with spin were studied previously e.g.
in~\cite{bol2,bol1,bul,gna}, but the attention was mostly concentrated
on Dirac- and Pauli-type operators. The theory of Rashba Hamiltonians is not developed
even in the Euclidian spaces, where the spin-orbit interaction promises
to show effects which are absent in the scalar case, like embedded
eigenvalues in short-range potentials or localization in crystals,
giving hence possibilities for constructing new nanodevices~\cite{MC,MC2}.

The aim of the paper is two-fold. First, we are going
to describe the Schr\"odinger operators
in two-dimensional networks with magnetic field and spin-orbit interaction.
An essential part here is the reduction of the quantum graph Hamiltonian
to a certain discrete equation. In the scalar case, an analogous procedure
was done in~\cite{exdual} for the solutions of the stationary Schr\"odinger
equation and  recently in~\cite{kp} for the spectra. Note that there is another approach
to the relationship between the quantum graphs and tight-binding Hamiltonians
coming from some asymptotic considerations~\cite{mo1,mo2}.
Second, by considering the localization effects
in the $T_3$-lattice we would like to attract the attention
of researchers working on quantum graphs to potential
applications in the study of superconducting networks.
In section~\ref{sec1} we give a mathematical formalism of quantum graphs
with external interactions; essentially we describe rigorously the constructions of~the works~\cite{r1,r2}. In section~\ref{sec2} we use this machinery to study the spectrum
of the $T_3$-lattice with a magnetic field and the Rashba interaction.
We show that the spectral problem is of supersymmetric type
and that the study of some energy levels is equivalent to the study of zero modes
in a certain discrete model.
As a result, we give a rigorous justification of the extreme localization
for the case of non-trivial scalar potentials on the edges and non-ideal
couplings at the nodes. We show that
at zero spin-orbit interaction this effect is independent of the edge potential.
At the same time, it appears that the generic Rashba interaction destroys
the localization. We also show that at a certain combination of the magnetic and spin parameters a new localization effect appears, where one can localize one of the spin
projections using the magnetic field.

\section{Quantum graphs with external interactions}\label{sec1}

\subsection{Schr\"odinger operator on a quantum graph embedded in Euclidian space}

In this section we describe the construction
of the Hamiltonian in a two-dimensional network with
magnetic field and Rashba interaction. Recall that the Rashba Hamiltonian
of a two-dimensional system acts on two-component vector functions as
takes the form~\cite{br2,ra}
\begin{equation}
      \label{eq-hso}
H=(\vcm p-\vcm A)^2+2\, k_R\, \langle \vcm p-\vcm A, \vc\sigma\times \vcm n\rangle+U,
\end{equation}
where $\vcm A$ is the magnetic vector potential, $k_R$ is the Rashba constant expressing the strength of the spin-orbit interaction,
$U$ is a scalar potential, $\vc \sigma$ is the vector of Pauli matrices,
and $\vcm n$ is the unit vector orthogonal to the plane of the system.
The second term,  which is the formal
mixed product, on the right-hand side of \eqref{eq-hso} takes into account
the spin-orbit coupling.
For $k_R=0$ the problem splits in two identical scalar problems.
The corresponding Hamiltonian for a network is obtained by projecting
all the interactions onto each edge and by introducing suitable boundary conditions
at the nodes, which will be described below. (We remark that some effects
of the Rashba interaction and the magnetic field in a wire can be studied
in other types of models~\cite{dk}.)

Let $\VV$ be a uniformly discrete subset of the $xy$-plane in $\RR^3$, the set of nodes (vertices).
The uniform discreteness means the existence of a constant $d>0$
such that $|\vc\alpha-\vc\beta|\ge d$ for all $\vc\alpha,\vc\beta\in \VV$ with $\vc\alpha\ne \vc\beta$.
Denote
\[
l_{\vc \alpha\vc\beta}:=|\vc\alpha-\vc\beta|,\quad
\vcm e_{\vc\alpha\vc\beta}:=\dfrac{1}{l_{\vc\alpha\vc\beta}}\big(\vc\beta-\vc\alpha\big).
\] 
Some nodes are connected by a directed edge. The set of all edges
will be denoted by $\EE$, $\EE\subset \VV\times \VV$.
The edge with initial vertex $\vc\alpha\in \VV$ and terminal vertex $\vc\beta\in \VV$
will be denoted by $\vc\alpha\vc\beta$. For
$\vc\alpha\in\VV$ denote $\indeg\vc\alpha:=\#\{\vc\beta\vc\alpha\in\EE\}$,
$\outdeg\vc\alpha=\#\{\vc\alpha\vc\beta\in\EE\}$,
$\deg\vc\alpha:=\indeg\vc\alpha+\outdeg\vc\alpha$. We assume that the degrees
satisfy the following conditions:
that
\begin{equation}
     \label{eq-deg}
\text{ there exists } N\in\ZZ \text{ with } 1\le\deg\vc\alpha\le N \text{ for all } \vc\alpha\in\VV;
\end{equation}
in particular, we assume that there are no isolated vertices.
The configuration consisting of all segments $[\vc\alpha,\vc\beta]$, $\vc\alpha\vc\beta\in\EE$
will be referred to as a metric graph or as a (wire) network. We assume that the system
has no self-intersections and that
\begin{equation}
      \label{eq-linf}
0<\inf\{l_{\vc\alpha\vc\beta}\}\le \sup\{l_{\vc\alpha\vc\beta}\}<\infty.
\end{equation}

The quantum state space corresponding to the metric graph is introduced as follows. 
Each edge $\vc\alpha\vc\beta$ will be identified with the segment
$[0,l_{\vc\alpha\vc\beta}]$ such that $\vc\alpha$ is identified
with $0$ and $\vc\beta$ is identified with $l_{\vc\alpha\vc\beta}$.
The state space of each edge $\vc\alpha\vc\beta$ is $\HH_{\vc\alpha\vc\beta}:=L^2\big([0,l_{\vc\alpha\vc\beta}],\CC^2)$.
The state space of the whole structure is
$\HH=\bigoplus_{\vc\alpha\vc\beta\in \EE} \HH_{\vc\alpha\vc\beta}$.

On each edge consider a real-valued scalar potential $U_{\vc\alpha\vc\beta}\in L^2[0,l_{\vc\alpha\vc\beta}]$.
To avoid unnecessary technical difficulties we will assume that the scalar potentials are uniformly $L^2$-bounded,
\begin{equation}
     \label{eq-usup}
\sup\|U_{\vc\alpha\vc\beta}\|_{L^2}<\infty.
\end{equation}
Assume that the system is subjected to an external magnetic field
given by a vector potential $\vcm A\in C^1(\RR^3,\RR^3)$. This induces
magnetic potentials on each edge,
$a_{\vc\alpha\vc\beta}(t):=\big\langle \vcm A(\vc \alpha+ t \vcm e_{\vc\alpha\vc\beta}),\vcm e_{\vc \alpha\vc\beta}\big\rangle$.

Denote by $k_R$ the Rashba constant. The spin-orbit interaction can be taken into account
by adding the term $2k_R \Big(-\bi\dfrac{\bd}{\bd t}-a_{\vc\alpha\vc\beta}(t)\Big) \langle \vc \sigma\times \vcm n, \vcm e_{\vc\alpha\vc\beta}\rangle$
with $\vcm n=(0,0,1)$. Therefore, the dynamics along each edge $\vc\alpha\vc\beta$
is described by the differential expression
\begin{multline*}
L_{\vc\alpha\vc\beta}=\Big(-\bi\dfrac{\bd}{\bd t}-a_{\vc\alpha\vc\beta}(t)\Big)^2
+2k_R \Big(-\bi\dfrac{\bd}{\bd t}-a_{\vc\alpha\vc\beta}(t)\Big) \langle\vc\sigma\times \vcm n,\vcm e_{\vc\alpha\vc\beta}\rangle+U_{\vc\alpha\vc\beta}\\
{}\equiv
\Big(\bi\dfrac{\bd}{\bd t}+a_{\vc\alpha\vc\beta}(t)+k_R\sigma_{\vc\alpha\vc\beta}\Big)^2+U_{\vc\alpha\vc\beta}-k_R^2,
\end{multline*}
where
\[
\sigma_{\vc\alpha\vc\beta}=\begin{pmatrix}
0 & e_{\vc\alpha\vc\beta2}+\bi e_{\vc\alpha\vc\beta1}\\
e_{\vc\alpha\vc\beta2}-\bi e_{\vc\alpha\vc\beta1} & 0
\end{pmatrix}.
\]
For a uniform magnetic field with the strength $\vcm B\in\RR^3$ it is useful
to use the symmetric gauge, $\vcm A(\vcm r)=\dfrac 12 \,\vcm B\times \vcm r$. In this
case the magnetic potentials $a_{\vc\alpha\vc\beta}$ are constant,
$a_{\vc\alpha\vc\beta}=\dfrac12 \langle \vcm B\times\vc\alpha,\vcm e_{\vc\alpha\vc\beta}\rangle$.

Denote by $L$ an operator in $\HH$
acting as
\begin{equation}
     \label{eq-th}
(\vcm f_{\vc\alpha\vc\beta})\mapsto (L_{\vc\alpha\vc\beta}\vcm f_{\vc\alpha\vc\beta})
\end{equation}
on functions $\vcm f_{\vc\alpha\vc\beta}\in H^2\big([0,l_{\vc\alpha\vc\beta}],\CC^2\big)$
satisfying at each $\vc\alpha\in \VV$:
\begin{subequations}
            \label{eq-fab}
\begin{gather}
            \label{eq-fabA}
\vcm f_{\vc \alpha\vc\beta}(0)=\vcm f_{\vc\gamma\vc\alpha}(l_{\vc\gamma\vc\alpha})=:\vcm f(\vc\alpha), \quad \vc\alpha\vc\beta,\, \vc\gamma\vc\alpha\in \EE,\\
            \label{eq-fabB}
\begin{gathered}
\sum_{\vc\alpha\vc\beta\in \EE} \Big[\dfrac{\bd}{\bd t}-\bi(a_{\vc\alpha\vc\beta}+k_R\sigma_{\vc\alpha\vc\beta})\Big]\vcm f_{\vc\alpha\vc\beta}(0)\\
{}-\sum_{\vc\beta\vc\alpha\in \EE} \Big[\dfrac{\bd}{\bd t}-\bi(a_{\vc\beta\vc\alpha}+k_R\sigma_{\vc\beta\vc\alpha})\Big]\vcm f_{\vc\beta\vc\alpha}(l_{\vc\beta\vc\alpha})=\epsilon(\vc\alpha) \vcm f(\alpha),
\end{gathered}
\end{gather}
\end{subequations}
where $\epsilon(\vc\alpha)$ are real-valued parameters. The case $\epsilon(\vc\alpha)=0$
may be considered as an ideal coupling, which is the analogue of the Kirchhoff coupling
in the scalar case. We are going to consider $L$ as the Hamiltonian of the system,
and our next aim is to show its self-adjointness.

\subsection{Self-adjointness and spectral analysis}

Denote by $\DD$ the set of all functions $\vcm f=(\vcm f_{\vc\alpha\vc\beta})$,
with components $\big(\vcm f_{\vc\alpha\vc\beta}\big)\in\bigoplus H^2\big([0,l_{\vc\alpha\vc\beta}],\CC^2\big)$,
$\vc\alpha\vc\beta\in\EE$, which are continuous at all nodes, i.e. such that
the condition~\eqref{eq-fabA} is satisfied. Clearly, for $\vcm f\in \DD$
the values $\vcm f(\vc\alpha)$, $\vc\alpha\in\VV$, have the direct sense.
Furthermore, for $\vcm f\in\DD$ and $\vc\alpha\in\VV$ denote
\begin{multline*}
\vcm f\,'(\vc\alpha):=\sum_{\vc\alpha\vc\beta\in \EE} \Big[\dfrac{\bd}{\bd t}-\bi(a_{\vc\alpha\vc\beta}+k_R\sigma_{\vc\alpha\vc\beta})\Big]\vcm f_{\vc\alpha\vc\beta}(0)\\
{}-\sum_{\vc\beta\vc\alpha\in \EE} \Big[\dfrac{\bd}{\bd t}-\bi(a_{\vc\beta\vc\alpha}+k_R\sigma_{\vc\beta\vc\alpha})\Big]\vcm f_{\vc\beta\vc\alpha}(l_{\vc\beta\vc\alpha}).
\end{multline*}
Consider in $\HH$ a linear operator $\Pi$ with domain $\DD$ acting by the rule~\eqref{eq-th}.

\begin{prop}\label{prop1} The operator $\Pi$ is closed.
For any $\vcm f\in\dom \Pi\equiv\DD$ the vectors $\Gamma \vcm f:=\big(\vcm f(\vc \alpha)\big)$
and $\Gamma' \vcm f:=\big(\vcm f'(\vc\alpha)\big)$ belong to $\ell^2(\VV,\CC^2)$,
and the map $(\Gamma,\Gamma'):\dom L\to \ell^2(\VV,\CC^2)\oplus \ell^2(\VV,\CC^2)$ is surjective.
For any $\vcm f,\vcm g\in\dom \Pi$ there holds
\begin{equation}
     \label{eq-part}
\langle \vcm f,\Pi\,\vcm g\rangle-\langle \Pi\,\vcm f,\vcm g\rangle=\langle\Gamma\vcm f,\Gamma'\vcm g\rangle-
\langle \Gamma'\vcm f,\Gamma \vcm g\rangle.
\end{equation}
\end{prop}

\begin{proof}
Denote by $\Theta_{\vc\alpha\vc\beta}$ the unitary transformation of $\HH_{\vc\alpha\vc\beta}$
given by 
\begin{equation}
    \label{eq-theta}
\Theta_{\vc\alpha\vc\beta}\,\vcm f(t)=\exp\Big(\bi\int_0^t \big(a_{\vc\alpha\vc\beta}(s)+k_R\sigma_{\vc\alpha\vc\beta}
\big)\bd s\Big) \vcm f(t).
\end{equation}
Denoting $\partial:=\dfrac{\bd}{\bd t}$ we see $(\partial-\bi a_{\vc\alpha\vc\beta}-\bi k_R\sigma_{\vc\alpha\vc\beta})\Theta_{\vc\alpha\vc\beta}\equiv \Theta_{\vc\alpha\vc\beta}\partial$.

By the Sobolev inequality, for any $c_1>0$ there exists
$c_2>0$ such that  for any $l>0$ and $\varphi\in H^2[0,l]$ there holds
\begin{gather*}
\|\varphi\|_\infty \le c_1\, l^{3/2}\,\|\varphi''\|_{L^2[0,l]}+
\dfrac{c_2}{l^{1/2}}\,\|\varphi\|_{L^2[0,l]},\\
\|\varphi'\|_\infty \le c_1\,l^{1/2}\,\|\varphi''\|_{L^2[0,l]}+
\dfrac{c_2}{l^{1/2}}\,\|\varphi\|_{L^2[0,l]}.
\end{gather*}
Note that for any $t\in[0,l_{\vc\alpha\vc\beta}]$ one has $\|\vcm f_{\vc\alpha\vc\beta}(t)\|_{\CC^2}
=\|\Theta_{\vc\alpha\vc\beta}\vcm f_{\vc\alpha\vc\beta}(t)\|_{\CC^2}$. Therefore, using the above estimate,
for any $\vcm f_{\vc\alpha\vc\beta}\in H^2\big([0,l_{\vc\alpha\vc\beta}],\CC^2\big)$ one has
\begin{align*}
\big\|\vcm f_{\vc\alpha\vc\beta}(t)\big\|_{\CC^2}&=\big\| \Theta^*_{\vc\alpha\vc\beta}\vcm f_{\vc\alpha\vc\beta} (t)\big\|_{\CC^2}\\
&\le c_1 l^{3/2}_{\vc\alpha\vc\beta}\big\| \partial^2 \Theta^*_{\vc\alpha\vc\beta}\vcm f_{\vc\alpha\vc\beta}\big\|_{\HH_{\vc\alpha\vc\beta}}
+ \dfrac{c_2}{l^{1/2}_{\vc\alpha\vc\beta}}\big\|\Theta^*_{\vc\alpha\vc\beta}\vcm f_{\vc\alpha\vc\beta}\big\|_{\HH_{\vc\alpha\vc\beta}}\\
&=c_1 l^{3/2}_{\vc\alpha\vc\beta}\big\| \Theta^*_{\vc\alpha\vc\beta}(\partial-\bi a_{\vc\alpha\vc\beta}-\bi k_R\sigma_{\vc\alpha\vc\beta})^2 \vcm f_{\vc\alpha\vc\beta}\big\|_{\HH_{\vc\alpha\vc\beta}}
+ \dfrac{c_2}{l^{1/2}_{\vc\alpha\vc\beta}}\big\|\Theta^*_{\vc\alpha\vc\beta}\vcm f_{\vc\alpha\vc\beta}\big\|_{\HH_{\vc\alpha\vc\beta}}\\
&=c_1 l_{\vc\alpha\vc\beta}^{3/2}\big\| (\partial-\bi a_{\vc\alpha\vc\beta}-\bi k_R\sigma_{\vc\alpha\vc\beta})^2 \vcm f_{\vc\alpha\vc\beta}\big\|_{\HH_{\vc\alpha\vc\beta}}
+ \dfrac{c_2}{l_{\vc\alpha\vc\beta}^{1/2}}\big\|\vcm f_{\vc\alpha\vc\beta}\big\|_{\HH_{\vc\alpha\vc\beta}}
\end{align*}
and, in the same way,
\begin{multline*}
\big\|(\partial-\bi a_{\vc\alpha\vc\beta}-\bi k_R\sigma_{\vc\alpha\vc\beta})\vcm f_{\vc\alpha\vc\beta}(t)\big\|_{\CC^2}\\
\le
c_1 l^{1/2}_{\vc\alpha\vc\beta}\big\| (\partial-\bi a_{\vc\alpha\vc\beta}-\bi k_R\sigma_{\vc\alpha\vc\beta})^2 \vcm f_{\vc\alpha\vc\beta}\big\|_{\HH_{\vc\alpha\vc\beta}}
+ \dfrac{c_2}{l_{\vc\alpha\vc\beta}^{1/2}}\big\|\vcm f_{\vc\alpha\vc\beta}\big\|_{\HH_{\vc\alpha\vc\beta}}. 
\end{multline*}
Using the assumptions~\eqref{eq-linf} and~\eqref{eq-usup} we conclude that
there exist positive constants $C_1$ and $C_2$ such that for any $\vc\alpha\vc\beta\in\EE$,
$\vcm f_{\vc\alpha\vc\beta}\in H^2\big([0,l_{\vc\alpha\vc\beta}],\CC^2\big)$, $t\in [0,l_{\vc\alpha\vc\beta}]$ one has
\begin{subequations}
    \label{eq-bdp}
\begin{gather}
     \label{eq-bdp1}
\big\|\vcm f_{\vc\alpha\vc\beta}(t)\big\|
\le C_1\|L_{\vc\alpha\vc\beta}f_{\vc\alpha\vc\beta}\|
+C_2\|f_{\vc\alpha\vc\beta}\|,\\
     \label{eq-bdp2}
\big\|(\partial-\bi a_{\vc\alpha\vc\beta}-\bi k_R\sigma_{\vc\alpha\vc\beta})\vcm f_{\vc\alpha\vc\beta}(t)\big\|
\le C_1\|L_{\vc\alpha\vc\beta}f_{\vc\alpha\vc\beta}\|
+C_2\|f_{\vc\alpha\vc\beta}\|.
\end{gather}
\end{subequations}
Here the norms  are taken in $\CC^2$ on the left-hand side
and  in $\HH_{\vc\alpha\vc\beta}$ on the right-hand side.

Denote by  $\widetilde \Pi$ the operator acting in $\HH$ by the rule~\eqref{eq-th}
on the domain $\dom\widetilde\Pi=\bigoplus_{\vc\alpha\vc\beta\in\EE} H^2\big([0,l_{\vc\alpha\vc\beta}],\CC^2\big)$.
Clearly, $\widetilde \Pi$ is closed. By~\eqref{eq-bdp1}, the linear maps
\[
T_{\vc\alpha\vc\beta\vc\gamma}:\dom \widetilde\Pi\ni
\vcm f\mapsto \vcm f_{\vc\alpha\vc\beta}(0)-\vcm f_{\vc\gamma\vc\alpha}(l_{\vc\gamma\vc\alpha})\in\CC^2,\quad
\vc\alpha\vc\beta, \vc\gamma\vc\alpha\in \EE,
\]
 are bounded with respect to the graph norm of $\widetilde\Pi$.
Therefore, the restriction of $\widetilde\Pi$ to the subspace where all these functionals
vanish is a closed operator. As this restriction is exactly $\Pi$, the operator
$\Pi$ is closed.

For $\vcm f\in\DD$ the inclusions $\Gamma \vcm f$, $\Gamma' \vcm f \in \ell^2(\VV,\CC^2)$ follow
immediately from the estimates~\eqref{eq-bdp}
and the assumption~\eqref{eq-deg}, and the identity~\eqref{eq-part}
can be verified directly using the partial integration. 

To prove the surjectivity condition, we fix first four functions $f_{jk}\in H^2[0,1]$ with $f_{jk}^{(i)}(l)=\delta_{ij}\delta_{kl}$, $i,j,k,l\in\{0,1\}$. Take arbitrary $\vc\xi,\vc\xi'\in \ell^2(\VV,\CC^2)$.
Denote
\begin{multline}
    \label{eq-tau}
\tau_{\vc\alpha\vc\beta}:=\exp\Big(\bi\int_0^{l_{\vc\alpha\vc\beta}} \big(a_{\vc\alpha\vc\beta}(s)+k_R\sigma_{\vc\alpha\vc\beta}\big)\bd s\Big)\\
\equiv
\exp\Big(\bi\int_0^{l_{\vc\alpha\vc\beta}} a_{\vc\alpha\vc\beta}(s)\bd s\Big)
\Big(\cos k_R \,l_{\vc\alpha\vc\beta}+\bi\,\sigma_{\vc\alpha\vc\beta}\,\sin k_R\, l_{\vc\alpha\vc\beta}\Big)
\in \mathbf{U}(2).
\end{multline}
By direct calculation, the function $\vcm f\in\HH$ whose components are of the form
$\vcm f_{\vc\alpha\vc\beta}=\Theta_{\vc\alpha\vc\beta}\vcm g_{\vc\alpha\vc\beta}$, where
\begin{multline*}
\vcm g_{\vc\alpha\vc\beta}(t)=
f_{00}\Big(\dfrac{t}{l_{\vc\alpha\vc\beta}}\Big)\vc\xi(\vc\alpha)
+
f_{01}\Big(\dfrac{t}{l_{\vc\alpha\vc\beta}}\Big)\tau^*_{\vc\alpha\vc\beta}\vc\xi(\vc\beta)\\
+\dfrac{l_{\vc\alpha\vc\beta}}{\deg\vc\alpha}\,
f_{10}\Big(\dfrac{t}{l_{\vc\alpha\vc\beta}}\Big)\vc\xi'(\vc\alpha)
-\dfrac{l_{\vc\alpha\vc\beta}}{\deg\vc\beta}\,
f_{11}\Big(\dfrac{t}{l_{\vc\alpha\vc\beta}}\Big)\tau^*_{\vc\alpha\vc\beta}\vc\xi'(\vc\beta),
\end{multline*}
lies in $\DD$ and satisfies $(\Gamma\vcm f,\Gamma'\vcm f)=(\vc\xi,\vc\xi')$.
\end{proof}

Proposition~\ref{prop1} shows that the space $\GG:=\ell^2(\VV,\CC^2)$ and the maps $\Gamma,\Gamma':\dom \Pi\to \GG$
form a \emph{boundary triple} for $\Pi$, see e.g.~\cite{dm,pavlov} for a detailed discussion. The self-adjointness
of $\Pi$ would follow from the following assertion~\cite{dm}: if $\Pi$ has at least one self-adjoint restriction
(i.e. if $\Pi^*$ is symmetric)
and $A$ is a self-adjoint operator in $\GG$,
then the restriction of $\Pi$ to the vectors $\varphi\in\dom\Pi$ satisfying
$\Gamma'\varphi=A\Gamma\varphi$ is self-adjoint in $\HH$.

Consider the restriction $D$ of $\Pi$ to the functions $\vcm f$ satisfying $\Gamma\vcm f=\vcm 0$.
Clearly, this restriction is nothing but the direct sum $\bigoplus_{\vc\alpha\vc\beta\in\EE} D_{\vc\alpha\vc\beta}$,
where $D_{\vc\alpha\vc\beta}$ is an operator in $\HH_{\vc\alpha\vc\beta}$
acting as $\vcm f_{\vc\alpha\vc\beta}\mapsto L_{\vc\alpha\vc\beta} \vcm f_{\vc\alpha\vc\beta}$
on functions satisfying $\vcm f_{\vc\alpha\vc\beta}(0)=\vcm f_{\vc\alpha\vc\beta}(l_{\vc\alpha\vc\beta})=\vcm0$.
As each $D_{\vc\alpha\vc\beta}$ is self-adjoint, so is $D$.
Note that  $L$ itself is the restriction of $\Pi$ to the functions
$\vcm f$ satisfying $\Gamma'\vcm f=T \Gamma \vcm f$, $T=\diag\big(\epsilon(\vc\alpha)\big)$.
This implies
\begin{prop}
The spin-orbit Hamiltonian $L$ is self-adjoint.
\end{prop}
To carry out the spectral analysis of $L$ it is useful to relate the resolvents of $L$ and $D$
by Krein's resolvent formula~\cite{dm},
\begin{equation}
         \label{eq-krein}
(D-E)^{-1}-(L-E)^{-1}=\gamma(E)\big[M(E)-T\big]^{-1}\gamma\,^*(\overline E)
\end{equation}
where $E\notin\spec L\cup\spec D$ and the maps
$\gamma(E)$ and $M(E)$ are defined as follows.
For a given $E\notin\spec D$ and $\vc\xi\in \ell^2(\VV,\CC^2)$, the function
$\gamma(E)\vc\xi=(\vcm f_{\vc\alpha\vc\beta})$ is the solution to $(\Pi-E)\vcm f=\vcm 0$
satisfying $\Gamma \vcm f=\vc\xi$.
The map $M(E):\ell^2(\VV,\CC^2)\to\ell^2(\VV,\CC^2)$ is given by $M(E)=\Gamma'\gamma(E)$.
A direct consequence of Eq. \eqref{eq-krein} is the relationship
\begin{equation}
   \label{eq-spec}
\spec L\setminus\spec D =\Big\{E\notin\spec D:\,0\in\spec\big(M(E)-T\big)\Big\}.
\end{equation}
Moreover, $E\notin\spec D$ is an eigenvalue of $L$ iff $0$ is an eigenvalue of $M(E)-T$,
and $\gamma(E)\ker\big(M(E)-T\big)$ is the corresponding eigensubspace~\cite{gm}.

Denote by $\widetilde D_{\vc\alpha\vc\beta}$ the self-adjoint operator in $L^2[0,l_{\vc\alpha\vc\beta}]$
acting as $g_{\vc \alpha\vc\beta}\mapsto -g''_{\vc \alpha\vc\beta}+(U_{\vc\alpha\vc\beta}-k_R^2) g_{\vc\alpha\vc\beta}$
on functions $g_{\vc \alpha\vc\beta}\in H^2[0,l_{\vc \alpha\vc\beta}]$ satisfying
the Dirichlet boundary condition, $g_{\vc \alpha\vc\beta}(0)=g_{\vc \alpha\vc\beta}(l_{\vc\alpha\vc\beta})$.
Note that the operators $\Theta^*_{\vc\alpha\vc\beta}D_{\vc\alpha\vc\beta}\Theta_{\vc\alpha\vc\beta}$
are of the form $\widetilde D_{\vc\alpha\vc\beta}\oplus\widetilde D_{\vc\alpha\vc\beta}$. In particular,
the spectra of $D_{\vc\alpha\vc\beta}$ coincide with those of $\widetilde D_{\vc\alpha\vc\beta}$ and are discrete sets, and
$\spec D=\overline{\bigcup_{\vc\alpha\vc\beta\in\EE} \spec 
\widetilde D_{\vc\alpha\vc\beta} }$.
 
Eq. \eqref{eq-spec} shows that the spectrum of $L$ outside $\spec D$ is completely described in terms of  $M(E)$. The question
whether $\spec D$ or some parts of it enter to the spectrum of $L$
must be analyzed individually taking into account the magnetic and spin parameters
and the topological properties of the graph.

Therefore, to carry out the spectral analysis for $L$ it is useful to calculate
the map $M(E)$. This can be done in terms of special (scalar) solutions to the equation
\begin{equation}
      \label{eq-sc}
-y''+U_{\vc\alpha\vc\beta} y= zy,\quad z\in\CC.
\end{equation}
Namely, denote by $s_{\vc\alpha\vc\beta}$ and $c_{\vc\alpha\vc\beta}$ the uniquely determined
solutions of \eqref{eq-sc} satisfying the boundary conditions
\[
s_{\vc\alpha\vc\beta}(0;z)=c'_{\vc\alpha\vc\beta}(0;z)=0,\quad
s'_{\vc\alpha\vc\beta}(0;z)=c_{\vc\alpha\vc\beta}(0;z)=1.
\]
Now let $\vc\xi\in \ell^2(\VV,\CC^2)$. To find $\gamma(E)\vc\xi=:(\vcm f_{\vc\alpha\vc\beta})$
we need to solve the boundary value problems
\begin{equation}
    \label{eq-bpg}
\begin{gathered}
\Big[\big(\bi \dfrac{\bd}{\bd t}+ a_{\vc\alpha\vc\beta}+k_R\sigma_{\vc\alpha\vc\beta}\big)^2
+U_{\vc\alpha\vc\beta}-k_R^2\Big]\vcm f_{\vc\alpha\vc\beta}= E \vcm f_{\vc\alpha\vc\beta},\\
\vcm f_{\vc\alpha\vc\beta}(0)=\vc\xi(\vc\alpha),
\quad
\vcm f_{\vc\alpha\vc\beta}(l_{\vc\alpha\vc\beta})=\vc\xi(\vc\beta).
\end{gathered}
\end{equation}
Writing $\vcm f_{\vc\alpha\vc\beta}:=\Theta_{\vc\alpha\vc\beta}\vcm g_{\vc\alpha\vc\beta}$,
where $\Theta_{\vc\alpha\vc\beta}$ is the unitary transformation from~\eqref{eq-theta}, we rewrite
\eqref{eq-bpg} as a boundary value problem for $\vcm g_{\vc\alpha\vc\beta}$,
\begin{equation}
    \label{eq-bpg2}
\begin{gathered}
-\vcm g''_{\vc\alpha\vc\beta}+ U_{\vc\alpha\vc\beta}\vcm g_{\vc\alpha\vc\beta}=(E+k_R^2) \vcm g_{\vc\alpha\vc\beta},\\
\vcm g_{\vc\alpha\vc\beta}(0)=\vc\xi(\vc\alpha),
\quad \vcm g_{\vc\alpha\vc\beta}(l_{\vc\alpha\vc\beta})=\tau_{\vc\alpha\vc\beta}^*\vc\xi(\vc\beta).
\end{gathered}
\end{equation}
The solution to \eqref{eq-bpg2} takes the form
\begin{multline*}
\vcm g_{\vc\alpha\vc\beta}(t)= \dfrac{s_{\vc\alpha\vc\beta}(t;E+k_R^2)}{s_{\vc\alpha\vc\beta}(l_{\vc\alpha\vc\beta};E+k_R^2)}\cdot
\Big[\tau^*_{\vc\alpha\vc\beta}\,\vc\xi(\vc\beta)-
c_{\vc\alpha\vc\beta}(l_{\vc\alpha\vc\beta};E+k_R^2)\, \vc\xi(\vc\alpha) \Big]\\
+c_{\vc\alpha\vc\beta}(t;E+k_R^2)\,\vc\xi(\vc\alpha).
\end{multline*}
Now we have 
\begin{gather}
\vcm g'_{\vc\alpha\vc\beta}(0)=\dfrac{1}{s_{\vc\alpha\vc\beta}(l_{\vc\alpha\vc\beta};E+k_R^2)}\Big[
\tau^*_{\vc\alpha\vc\beta}\vc\xi(\vc\beta)-c_{\vc\alpha\vc\beta}(l_{\vc\alpha\vc\beta};E+k_R^2)\vc\xi(\vc\alpha)
\Big],\\
\vcm g'_{\vc\alpha\vc\beta}(l_{\vc\alpha\vc\beta})=\dfrac{1}{s_{\vc\alpha\vc\beta}(l_{\vc\alpha\vc\beta};E+k_R^2)}
\Big[s'_{\vc\alpha\vc\beta}(l_{\vc\alpha\vc\beta};E+k_R^2)\,\tau^*_{\vc\alpha\vc\beta}\,
\vc\xi(\vc\beta)-
\vc\xi(\vc\alpha)\Big].
\end{gather}
Noting that
\[
\vcm f'(\vc\alpha)=\sum_{\vc\alpha\vc\beta\in\EE} \vcm g'_{\vc\alpha\vc\beta}
-\sum_{\vc\beta\vc\alpha\in\EE}\tau_{\vc\beta\vc\alpha} \vcm g'_{\vc\alpha\vc\beta}(l_{\vc\alpha\vc\beta})
\]
we arrive at
\begin{multline}
    \label{eq-Mz}
M(E)\vc\xi(\vc\alpha)=
\sum_{\vc\alpha\vc\beta\in \EE}\dfrac{1}{s_{\vc\alpha\vc\beta}(l_{\vc\alpha\vc\beta};E+k_R^2)}\,\tau^*_{\vc\alpha\vc\beta}\,\vc\xi(\vc\beta)\\
+
\sum_{\vc\beta\vc\alpha\in \EE}\dfrac{1}{s_{\vc\beta\vc\alpha}(l_{\vc\beta\vc\alpha};E+k_R^2)}\,\tau_{\vc\beta\vc\alpha}\,\vc\xi(\vc\beta)\\
-
\Big[
\sum_{\vc\alpha\vc\beta\in \EE} \dfrac{c_{\vc\alpha\vc\beta}(l_{\vc\alpha\vc\beta};E+k_R^2)}{s_{\vc\alpha\vc\beta}(l_{\vc\alpha\vc\beta};E+k_R^2)}
+
\sum_{\vc\beta\vc\alpha\in \EE} \dfrac{s'_{\vc\beta\vc\alpha}(l_{\vc\beta\vc\alpha};E+k_R^2)}{s_{\vc\beta\vc\alpha}(l_{\vc\beta\vc\alpha};E+k_R^2)}
\Big]\,\vc\xi(\vc\alpha).
\end{multline}
Using Krein's resolvent formula~\eqref{eq-krein} we come to
\begin{thm}\label{prop3} The set $\spec L\setminus\spec D$ consists exactly of the real numbers $E$
such that $0\in\spec\big[M(E)-T\big]$, where $M(E)$ and $T$ are operators in $\ell^2(\VV,\CC^2)$,
$M(E)$ is given by~\eqref{eq-Mz}
and $T=\diag\big(\epsilon(\vc\alpha)\big)$. Moreover, such $E$ is an eigenvalue of $L$ iff $0$ is an eigenvalue of $M(E)-T$,
and $\gamma(E)\ker\big(M(E)-T\big)$ is the corresponding eigenspace.
\end{thm}
We remark that in the above calculations it does not matter
whether $\vc\xi$ is in $\ell^2$ or not. Actually, all the construction hold for any set of vectors
$\vc \xi(\vc\alpha)\in\CC^2$, $\vc\alpha\in\VV$. This observation can be formulated as follows:
\begin{thm}
For $E\notin\spec D$, any continuous solution $\vcm f$ to $(L-E)\vcm f=0$ has the form
\begin{multline*}
\vcm f_{\vc\alpha\vc\beta}(t)= \dfrac{s_{\vc\alpha\vc\beta}(t;E+k_R^2)}{s_{\vc\alpha\vc\beta}(l_{\vc\alpha\vc\beta};E+k_R^2)}\cdot
\Theta_{\vc\alpha\vc\beta}\Big[\tau^*_{\vc\alpha\vc\beta}\,\vcm f(\vc\beta)-
c_{\vc\alpha\vc\beta}(l_{\vc\alpha\vc\beta};E+k_R^2)\, \vcm f(\vc\alpha) \Big]\\
+\Theta_{\vc\alpha\vc\beta}c_{\vc\alpha\vc\beta}(t;E+k_R^2)\,\vcm f(\vc\alpha).
\end{multline*}
Such a function satisfies the boundary conditions~\eqref{eq-fabB} iff
\begin{multline*}
\sum_{\vc\alpha\vc\beta\in \EE}\dfrac{1}{s_{\vc\alpha\vc\beta}(l_{\vc\alpha\vc\beta};E+k_R^2)}\,\tau^*_{\vc\alpha\vc\beta}\,\vcm f(\vc\beta)
+
\sum_{\vc\beta\vc\alpha\in \EE}\dfrac{1}{s_{\vc\beta\vc\alpha}(l_{\vc\beta\vc\alpha};E+k_R^2)}\,\tau_{\vc\beta\vc\alpha}\,\vcm f(\vc\beta)\\
=
\Big[
\sum_{\vc\alpha\vc\beta\in \EE} \dfrac{c_{\vc\alpha\vc\beta}(l_{\vc\alpha\vc\beta};E+k_R^2)}{s_{\vc\alpha\vc\beta}(l_{\vc\alpha\vc\beta};E+k_R^2)}
+
\sum_{\vc\beta\vc\alpha\in \EE} \dfrac{s'_{\vc\beta\vc\alpha}(l_{\vc\beta\vc\alpha};E+k_R^2)}{s_{\vc\beta\vc\alpha}(l_{\vc\beta\vc\alpha};E+k_R^2)}
+\epsilon(\vc\alpha) \Big]\,\vcm f(\vc\alpha).
\end{multline*}
Such an $\vcm f$ is an eigenfunction of $L$ (i.e. belongs to $L^2$) iff $\big(\vcm f(\vc\alpha)\big)_{\vc\alpha\in\VV}\in\ell^2(\VV,\CC^2)$.
\end{thm}
Note that similar formulas for more simple situations were obtained earlier e.g.in~\cite{akk,des,jvb,Ku1,exdual,r2}. 

The expression~\eqref{eq-Mz} can be simplified significantly if all the edges are the same,
i.e. if $l_{\vc\alpha\vc\beta}\equiv l$ and $U_{\vc\alpha\vc\beta}\equiv U$,
$s_{\vc\alpha\vc\beta}=s$, $c_{\vc\alpha\vc\beta}=c$ for all $\vc\alpha\vc\beta\in \EE$.
Note that in this case the spectrum of $D$ coincides with the Dirichlet spectrum of the operator
$\widetilde D:=-\bd^2/\bd t^2+U-k_R^2$ on the segment $[0,l]$ and hence is a discrete set.
We have
\begin{multline}
   \label{eq-meq}
M(E)\vc\xi(\vc\alpha)=
\dfrac{1}{s(l;E+k_R^2)}\bigg\{
\Big[
\sum_{\vc\alpha\vc\beta\in \EE}\tau^*_{\vc\alpha\vc\beta}\,\vc\xi(\vc\beta)
+
\sum_{\vc\beta\vc\alpha\in\EE}\tau_{\vc\beta\vc\alpha}\,\vc\xi(\vc\beta)
\Big]\\
-
\Big[\outdeg \vc\alpha\, c(l;E+k_R^2)
+
\indeg \vc\alpha \,s'(l;E+k_R^2)
\Big]
\,\vc\xi(\vc\alpha)\bigg\}.
\end{multline}
Even this expression admits further simplifications. 

\begin{prop}\label{prop5}
Assume that all edges are identical, $l_{\vc\alpha\vc\beta}\equiv l$,
$U_{\vc\alpha\vc\beta}\equiv U$, $U$ is even, $ U(t)\equiv U(l-x)$,
and the coupling constants $\epsilon(\vc\alpha)$
are of the form $\epsilon(\vc\alpha)=\deg\vc\alpha\,\epsilon$, then
$\spec L\setminus\spec\widetilde D=t_\epsilon^{-1}(\spec\Delta)$, where
$t_\epsilon(E)=c(l;E+k_R^2)+\epsilon s(l;E+k_R^2)$ and $\Delta$ is the discrete
Hamiltonian,
\[
\Delta \vc\xi(\vc\alpha)=\dfrac{1}{\deg\vc\alpha}\,\Big(\sum_{\vc\alpha\vc\beta\in \EE}\tau^*_{\vc\alpha\vc\beta}\,\vc\xi(\vc\beta)
+ \sum_{\vc\beta\vc\alpha\in\EE}\tau_{\vc\beta\vc\alpha}\,\vc\xi(\vc\beta)\Big),
\]
acting on the space $\ell^2(\VV,\CC^2;\deg)$ with the scalar product
\[
\langle \vc\xi,\vc\eta\rangle_{\deg}=\sum_{\vc\alpha\in\VV}
\deg\vc\alpha\cdot\overline{\vc\xi(\vc\alpha)}\,\vc\eta(\vc\alpha).
\]
\end{prop}

\begin{proof}
If the potential $U$ is even, one has $c(l;E+k_R^2)\equiv s'(l;E+k_R^2):=t(E)$,
see e.g.~\cite{kp}, hence
\[
M(E)-T=\dfrac{1}{s(l;E+k_R^2)}\Big[
\widetilde\Delta- t_\epsilon(E)\,\deg\Big],\quad \deg=\diag\big(\deg\vc\alpha\big),
\]
where $\widetilde \Delta$ is the discrete Hamiltonian in $\ell^2(\VV,\CC^2)$,
\[
\widetilde \Delta \vc\xi(\vc\alpha)=\sum_{\vc\alpha\vc\beta\in \EE}\tau^*_{\vc\alpha\vc\beta}\,\vc\xi(\vc\beta)
+ \sum_{\vc\beta\vc\alpha\in\EE}\tau_{\vc\beta\vc\alpha}\,\vc\xi(\vc\beta).
\]
The condition $0\in\spec\big[M(E)-T\big]$ takes the form
$0\in\spec\big[\widetilde\Delta-t_\epsilon(E)\deg\big]$ in $\ell^2(\VV,\CC^2)$,
which is equivalent to $0\in\spec\big[\Delta-t_\epsilon(E)\big]$ in $\ell^2(\VV,\CC^2;\deg)$.
\end{proof}

Proposition~\ref{prop5} shows that the spectral problem for a class of quantum graphs
reduces to the study of the tight-binding Hamiltonian $\Delta$. In the case $U\equiv 0$
and $\epsilon(\vc\alpha)\equiv0$ one has $t_\epsilon(E)=\cos\sqrt{E+k_R^2}$, and we arrive at
$\spec L=\Arccos^2\spec \Delta-k_R^2$ (up to the discrete set $\spec\widetilde D$), which
is exactly the formula connecting the network and the tight-binding spectra in the de Gennes-Alexander model of superconductivity~\cite{alx}. For scalar situation,
an analogue of this correspondence
was given e.g. in~\cite{jvb} for the Laplacian on compact graphs,
in~\cite{CC} for the Laplacian on non-compact graphs, and in~\cite{kp}
for more general Schr\"odinger operators.
At the same time, proposition~\ref{prop5} does not exhaust all possibilities
of such a reduction, i.e. the reduction to a discrete Hamiltonian is possible also for some non-even $U$.
(Such questions will be  discussed in greater detail in~\cite{kp2}.)
One of such situations will be discussed in the next section.

\section{Spectrum of $T_3$-lattice}\label{sec2}

\subsection{Description of the lattice}

\begin{figure}
\centering
\includegraphics[width=60mm]{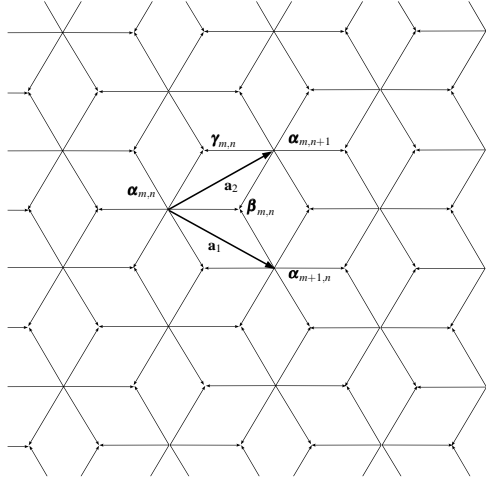}
\caption{A finite piece of $T_3$-lattice}
          \label{fig-t3}
\end{figure}

In this section, we consider the spectral problem for a quantum graph
whose underlying structure is the so-called $T_3$-lattice (see figure~\ref{fig-t3}).
The nodes are the points $\vc\alpha_{m,n}$, $\vc\beta_{m,n}$, $\vc\gamma_{m,n}$
with $\vc\alpha_{m,n}=m\vcm a_1+n\vcm a_2$, $\vcm a_1=(\dfrac32,-\dfrac{\sqrt{3}}{2},0)$,
$\vcm a_2=(\dfrac32,\dfrac{\sqrt{3}}{2},0)$, $\vc\beta_{m,n}=\vc\alpha_{m,n}+(1,0,0)$,
$\vc\gamma_{m,n}=\vc\alpha_{m,n}+(\dfrac12,\dfrac{\sqrt{3}}{2},0)$, $m,n\in\ZZ$,
i.e.
\begin{gather*}
\vc\alpha_{m,n}=\Big(\frac{3(m+n)}{2},\frac{\sqrt{3}(n-m)}{2},0\Big),\quad
\vc\beta_{m,n}=\Big(\frac{3(m+n)+2}{2},\frac{\sqrt{3}(n-m)}{2},0\Big),\\
\vc\gamma_{m,n}=\Big(\frac{3(m+n)+1}{2},\frac{\sqrt{3}(n-m+1)}{2},0\Big).
\end{gather*}
The edges are
\begin{align*}
e_{m,n,1}&=\vc\alpha_{m,n}\vc\gamma_{m,n},&
e_{m,n,2}&=\vc\alpha_{m,n}\vc\beta_{m-1,n},&
e_{m,n,3}&=\vc\alpha_{m,n}\vc\gamma_{m,n-1},\\
e_{m,n,4}&=\vc\alpha_{m,n}\vc\beta_{m,n-1},&
e_{m,n,5}&=\vc\alpha_{m,n}\vc\gamma_{m+1,n-1},&
e_{m,n,6}&=\vc\alpha_{m,n}\vc\beta_{m,n}.
\end{align*}
All the edges have the length $1$.
The direction vectors of $e_{m,n,j}$ are $\vcm e_j=\big(\cos\dfrac{\pi j}{3},\sin\dfrac{\pi j}{3},0\big)$, $j=1,\dots,6$.

\subsection{Reduction to tight-binding Hamiltonian}

We will assume that the system is subjected to the following external interactions.
On each edge there is the same  potential $U\in L^2[0,1]$.
The lattice is subjected to the uniform magnetic field $\vcm B=(0,0,2\pi\xi)$
orthogonal to the plane, and the magnetic vector potential in the symmetric gauge is $\vcm A(\vcm x)=(-\pi\xi x_2,\pi\xi x_1,0)$.
In what follows we use the magnetic parameter $\omega=\dfrac{\pi \xi\sqrt{3}}{2}$
expressing the magnetic flux through the elementary rhombus
(for example, $\vc\alpha_{m,n}\vc\beta_{m,n}\vc\alpha_{m,n+1}\vc\gamma_{mn}$).

The external magnetic field implies non-trivial magnetic potentials on $e_{m,n,j}$,
$a_{m,n,j}=\dfrac{1}{2} \langle \vcm B\times \vc\alpha_{m,n},\vcm e_j\rangle$,
\begin{align*}
a_{m,n,1}&=\omega (2m+n), & a_{m,n,2}&=\omega (m+2n), &
a_{m,n,3}&=\omega (n-m), \\ a_{m,n,4}&=-\omega (2m+n), &
a_{m,n,5}&=-\omega (m+2n), & a_{m,n,3}&=\omega (m-n).
\end{align*}
The dynamics along $e_{m,n,j}$ is described by the differential expression
\begin{gather*}
L_{m,n,j}=\big(\bi\dfrac{\bd}{\bd t}+a_{m,n,j}+k_R\,\sigma_{m,n,j}\big)^2+U-k_R^2,\\
\sigma_{m,n,j}=\begin{pmatrix}
0 & e_{j2}+ie_{j1}\\
e_{j2}-ie_{j1} &0
\end{pmatrix}
\equiv
\begin{pmatrix}
0 & \exp\Big[\bi\big(\dfrac{\pi}{2}-\dfrac{\pi j}{3}\big)\Big]\\
\exp\Big[\bi\big(\dfrac{\pi j}{3}-\dfrac{\pi}{2}\big)\Big] &0
\end{pmatrix},\\
m,n\in\ZZ,\quad j\in\{1,\dots,6\}.
\end{gather*}
Here $e_{jk}$ are the components of the vectors $\vcm e_j$
and $k_R$ is the Rashba constant.
We consider boundary conditions of the form \eqref{eq-fab}
at all the nodes
assuming that there are only two types of coupling constants:
$\lambda:=\epsilon(\vc\alpha_{m,n})$ and $\mu:=\epsilon(\vc\beta_{m,n})=\epsilon(\vc\gamma_{m,n})$.

The corresponding matrices $\tau_{m,n,j}$ from Eq.~\eqref{eq-tau},
$\tau_{m,n,j}=\exp\big[\bi(a_{m,n,j}+k_R\sigma_{m,n,j})\big]$, are as follows:
\begin{subequations}
\begin{align}
\tau_{m,n,1}=\tau^*_{m,n,4}&=\be^{\,\bi\omega(2m+n)}\Big\{
\cos k_R+\bi\sin k_R \begin{pmatrix}
0 & \be^{\,\bi\pi/6}\\
\be^{-\bi\pi/6} &0
\end{pmatrix}\Big\},\\
\tau_{m,n,2}=\tau^*_{m,n,5}&=\be^{\,\bi\omega(m+2n)}\Big\{
\cos k_R+\bi\sin k_R \begin{pmatrix}
0 & \be^{-\bi\pi/6}\\
\be^{\,\bi\pi/6} &0
\end{pmatrix}\Big\},\\
\tau_{m,n,3}=\tau^*_{m,n,6}&=\be^{\,\bi\omega(n-m)}\Big\{
\cos k_R+\bi\sin k_R \begin{pmatrix}
0 & \be^{-\bi\pi/2}\\
\be^{\,\bi\pi/2} &0
\end{pmatrix}\Big\}.
\end{align}
\end{subequations}
Clearly, for any $m,n\in\ZZ$ one has
\begin{gather*}
\outdeg \vc\alpha_{m,n}=6,\quad
\indeg \vc\beta_{m,n}=\indeg \vc\gamma_{m,n}=3,\\
\indeg \vc\alpha_{m,n}=\outdeg \vc\beta_{m,n}=\outdeg \vc\gamma_{m,n}=0.
\end{gather*}

For the subsequent analysis we use the fact that the lattice is bipartite.
Represent the set of nodes as the disjoint union
$\VV=\VV_0\cup \VV_1$, $\VV_0=\{\vc\alpha_{m,n}\}$, $\VV_1=\{\vc\beta_{m,n}\}\cup\{\vc\gamma_{m,n}\}$.
Clearly, for the set of edges one has $\EE\subset \VV_0\times \VV_1$.
With respect the  the decomposition $\ell^2(\VV,\CC^2)=\ell^2(\VV_0,\CC^2)\oplus \ell^2(\VV_1,\CC^2)$
the operator $T$ in theorem~\ref{prop3} takes the block-diagonal form,
\[
T=\begin{pmatrix}
\lambda & 0\\
0 & \mu
\end{pmatrix}.
\]
Using the above decomposition and 
Eq.~\eqref{eq-meq}, we rewrite $M(E)-T$ as
\begin{gather*}
M(E)-T=\dfrac{1}{s(1;E+k_R^2)}\,\begin{pmatrix}
-a(E) & A^*\\
A & -b(E)
\end{pmatrix},\\
a(E)=6c(1;E+k_R^2)+\lambda \,s(1;E+k_R^2),\\
b(E)=3s'(1;E+k_R^2)+\mu \,s(1;E+k_R^2).
\end{gather*}
where
\begin{align}
A\vcm f(\vc\beta_{m,n})=&
\tau_{m+1,n,2}\vcm f(\vc\alpha_{m+1,n})+\tau_{m,n+1,4}\vcm f(\vc\alpha_{m,n+1})+\tau_{m,n,6}\vcm f(\vc\alpha_{m,n}),\notag\\
A\vcm f(\vc\gamma_{m,n})=&
\tau_{m,n,1}\vcm f(\vc\alpha_{m,n})+\tau_{m,n+1,3}\vcm f(\vc\alpha_{m,n+1})+\tau_{m-1,n+1,5}\vcm f(\vc\alpha_{m-1,n+1}),\notag\\
         \label{eq-dstar}
A^*\vcm f(\vc\alpha_{m,n})=&\tau^{*}_{m,n,1}\vcm f(\vc\gamma_{m,n})+\tau^{*}_{m,n,2}\vcm f(\vc\beta_{m-1,n})
 +\tau^{*}_{m,n,3}\vcm f(\vc\gamma_{m,n-1})\\
&+\tau^{*}_{m,n,4}\vcm f(\vc\beta_{m,n-1})
 +\tau^{*}_{m,n,5}\vcm f(\vc\gamma_{m+1,n-1})+\tau^{*}_{m,n,6}\vcm f(\vc\beta_{m,n}).\notag
\end{align}
The operator $A^*$, $A^*:\ell^2(\VV_1,\CC^2)\to\ell^2(\VV_0,\CC^2)$
is adjoint to $A$, $A:\ell^2(\VV_0,\CC^2)\to\ell^2(\VV_1,\CC^2)$.
Using theorem~\ref{prop3} we write the condition
$E\in\spec L$ or, equivalently, $0\in\spec \big(M(E)-T\big)$, as
\begin{gather}
     \label{eq-abs}
\dfrac{a(E)+b(E)}{2}\in\spec\bigg[
\dfrac{b(E)-a(E)}{2}\,\begin{pmatrix}
1 & 0\\
0 & -1
\end{pmatrix}
+\begin{pmatrix}
0 & A^*\\
A & 0
\end{pmatrix}
\bigg].
\end{gather}
Note that $E\notin\spec \widetilde D$ in all the above constructions,
where $\widetilde D$ is the Dirichlet realization of $-\bd^2/\bd t^2+U-k_R^2$
on $[0,1]$. The question whether $\spec\widetilde D$ is a part of $\spec L$ or not
admits a simple answer in our case.
\begin{lemma} For all $\omega$ and $k_R$ one has
$\spec \widetilde D\subset \spec L$.
\end{lemma}

\begin{proof}
Using the Schnol-type arguments, cf.~\cite{bgp,Ku2}, it is sufficient to show
that for each $E\in\spec\widetilde D$ the equation $L\vcm f=E\vcm f$
has a bounded solution $\vcm f$ satisfying the boundary conditions~\eqref{eq-fab}.
Let $E\in\widetilde D$ and $g$ be the corresponding
eigenfunction of $\widetilde D$. Choose any infinite path $\PP$ without intersection
on the graph and any non-zero vector $\vcm z\in\CC^2$. For a fixed $e\in\PP$
set $\vcm f_e:=\Theta_e\, g\vcm z$, where $\Theta_e$ is given by~\eqref{eq-theta}.
Now extend $\vcm f$ to the whole graph in such a way that (a) $\vcm f_e=\vcm 0$
for $e\notin\PP$ and (b) on each $b\in\PP$ one has $\vcm f_b=\Theta_b\, g \vcm z_b$,
where the vectors $\vcm z_b$ are chosen in such a way that the boundary conditions~\eqref{eq-fab} are satisfied. By construction, there holds $L\vcm f=E\vcm f$.
At the same time, due to the unitarity of the matrices $\tau_{m,n,j}$ the obtained function $\vcm f$ is bounded. This finishes the proof.
\end{proof}

\subsection{Supersymmetric analysis}\label{sec3.3}
Eq. \eqref{eq-abs} is a typical supersymmetric spectral problem.
Using proposition~\ref{prop-susy} and corollary~\ref{corol-susy}
in Appendix one easily sees
that the set $\Sigma$ of $E$ for which the condition \eqref{eq-abs}
is satisfied is the union $\Sigma=\Sigma_1\cup\Sigma_2\cup\Sigma_3$,
\begin{gather*}
\Sigma_1=\Big\{
E\notin\spec \widetilde D: a(E)b(E)\ne 0 \text{ and }a(E)b(E)\in\spec A^*A,
\Big\}\\
\Sigma_2=\begin{cases}
\big\{
E\notin\spec \widetilde D:\,a(E)=0\big\}, & \text{if } 0\in\spec A^*A,\\
\emptyset, & \text{otherwise},
\end{cases}\\
\Sigma_3=\begin{cases}
\big\{
E\notin\spec  \widetilde D:\,b(E)=0\big\}, & \text{if } 0\in\spec AA^*,\\
\emptyset, & \text{otherwise}.
\end{cases}
\end{gather*}
To summarize,
\begin{prop}\label{prop-specL}
$\spec L=\Sigma_1\cup\Sigma_2\cup\Sigma_3\cup\spec\widetilde D$.
\end{prop}
Note that the sets $\Sigma_2$, $\Sigma_3$, and $\spec\widetilde D$ are discrete.
Therefore, only the set $\Sigma_1$ is responsible for the continuous
spectrum. Writing $\vcm f(\vc\alpha_{m,n})=:\vcm f(m,n)$,
we note that $A^*A$ is an operator
on $\ell^2(\ZZ^2,\CC^2)$ of the form
\begin{align*}
A^*A\vcm f(m,n)=&6\vcm f(m,n)\\
 &+  (\tau^{*}_{m,n,1}\tau_{m,n+1,3}+\tau^{*}_{m,n,6}\tau_{m,n+1,4})\vcm f(m,n+1)\\
 &+(\tau^{*}_{m,n,3}\tau_{m,n-1,1}+\tau^{*}_{m,n,4}\tau_{m,n-1,6})\vcm f(m,n-1)\\
 &+(\tau^{*}_{m,n,5}\tau_{m+1,n,3}+\tau^{*}_{m,n,6}\tau_{m+1,n,2})\vcm f(m+1,n)\\
 &+(\tau^{*}_{m,n,2}\tau_{m-1,n,6}+\tau^{*}_{m,n,3}\tau_{m-1,n,5})\vcm f(m-1,n)\\
 &+(\tau^{*}_{m,n,1}\tau_{m-1,n+1,5}+\tau^{*}_{m,n,2}\tau_{m-1,n+1,4})\vcm f(m-1,n+1)\\
 &+(\tau^{*}_{m,n,4}\tau_{m+1,n-1,2}+\tau^{*}_{m,n,5}\tau_{m+1,n-1,1})\vcm f(m+1,n-1),
\end{align*}
i.e.
\begin{multline}
A^*A =6 + \cos\omega\cdot\sin{2k_R}\cdot\widetilde\Delta\\
+2\,\bigg[\cos\omega\cdot\cos^2 k_R
-\sin^2k_R
\begin{pmatrix}
\cos\big(\omega-\dfrac{\pi}{3}\big) & 0\\
0 & \cos\big(\omega+\dfrac{\pi}{3}\big)
\end{pmatrix}
\bigg]\,
\begin{pmatrix}
\Delta & 0\\
0& \Delta
\end{pmatrix}
\end{multline}
where $\Delta$ is a spinless operator in $\ell^2(\ZZ^2)$,
\begin{align*}
\Delta f(m,n)&= \be^{-3\bi\omega m} f(m,n+1)+\be^{3\bi\omega m} f(m,n-1)\\
&+\be^{3\bi\omega n} f(m+1,n) +\be^{-3\bi\omega n} f(m-1,n)\\
&+\be^{-3\bi\omega(m+n)} f(m-1,n+1)+\be^{3\bi\omega(m+n)} f(m+1,n-1),
\end{align*}
and
\begin{align*}
\widetilde\Delta\vcm f(m,n)=& R_1\,
 \be^{-3\bi\omega m}\vcm f(m,n+1)+ R_1^*\,\be^{3\bi\omega m}\vcm f(m,n-1)\\
+& R_2\, \be^{3\bi\omega n}\vcm f(m+1,n)+ R_2^*\,\be^{-3\bi\omega n}\vcm f(m-1,n)\\
+& R_3\, \be^{-3\bi\omega(m+n)}\vcm f(m-1,n+1)+ R_3^*\,\be^{3\bi\omega(m+n)}\vcm f(m+1,n-1)
\end{align*}
with
\begin{gather*}
R_1=\begin{pmatrix}
0 & \dfrac32 -\dfrac{\sqrt 3}{2}\,\bi\\
-\dfrac32 -\dfrac{\sqrt 3}{2}\,\bi & 0
\end{pmatrix},
\quad
R_2=\begin{pmatrix}
0 & \dfrac32 +\dfrac{\sqrt 3}{2}\,\bi\\
-\dfrac32 +\dfrac{\sqrt 3}{2}\,\bi & 0
\end{pmatrix},\\
R_3=\begin{pmatrix}
0 & -\sqrt{3}\, \bi\\
-\sqrt{3}\, \bi & 0
\end{pmatrix}.
\end{gather*}
The expression for $A^*A$ shows explicitly the contribution of the magnetic and spin-orbit parameters
to the spectrum. Let us discuss the situations where the spectrum shows certain localization phenomena.

\subsection{Magnetic field induced extreme localization}\label{sec-loc-mag}

If the spin-orbit interaction is not taken into account,
$k_R=0$, then one has $A^*A=6+2\cos \omega\,(\Delta\oplus\Delta)$. In particular, at $\omega-\dfrac{\pi}{2}\in\pi\ZZ$
one has $A^*A=6$, i.e. the spectrum of $A^*A$ degenerates to a point.
If $\omega-\dfrac{\pi}{2}\in\pi\ZZ$ but the spin-orbit interaction is non-trivial,
similar phenomena occur only at certain values of the Rashba constant, i.e. $k_R\in\pi\ZZ$.
For generic values of $k_R$, obviously, there are some bands of continuous spectrum.

Let us analyze the sets $\Sigma_2$ and $\Sigma_3$ for this case, i.e. for $\omega-\dfrac{\pi}{2}\in\pi\ZZ$
and $k_R\in\pi\ZZ$. Clearly, the set $\Sigma_2$ is empty, as $0\notin\spec A^*A$. Let us look
at the operator $AA^*$.

\begin{lemma} For $\omega-\dfrac{\pi}{2}\in\pi\ZZ$
and $k_R\in\pi\ZZ$ one has $0\in\spec AA^*$.
\end{lemma}

\begin{figure}\centering
\includegraphics[width=60mm]{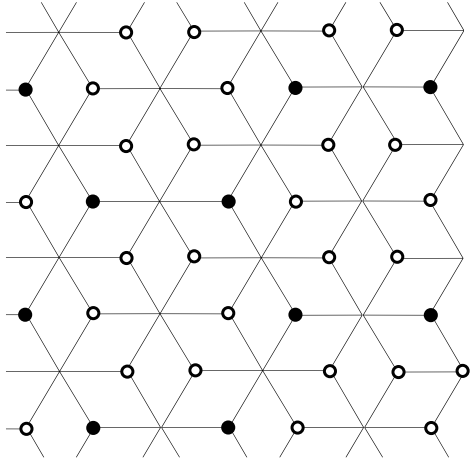}
\caption{Classification of the nodes $\vc \beta_{m,n}$ and $\vc\gamma_{m,n}$.}
         \label{fig-hex2}
\end{figure}

\begin{proof}
In view of periodicity, it is sufficient to show
that the equation $AA^*\psi=0$ has non-trivial bounded solutions,
$\psi\in \ell^\infty(\VV_1;\CC^2)$.

Let us classify the nodes $\vc\beta_{m,n}$
and $\vc\gamma_{m,n}$ as shown in figure~\ref{fig-hex2}. Consider all vector-valued
functions on $\VV_1$ vanishing at the white marked nodes. 
For such a function $\psi$, the condition $A^*\psi=0$ is of very simple form, because
in the expression~\eqref{eq-dstar} only two of the six terms on the right-hand side
are non-zero. Therefore, fixing the value of $\psi$ at a single
black marked node one uniquely extends $\psi$ to a bounded
solution of $A^*\psi=0$.
The conditions $\omega\in\dfrac{\pi}{2}+\pi\ZZ$ and $k_R\in\pi \ZZ$ guarantee
that this solution is well defined, i.e. that the phase factor along each cycle
on the hexagonal lattice of black nodes is $1$.
\end{proof}

To summarize the previous considerations we note that the set $\spec\widetilde D$
consists of the real $E$ satisfying $s(1;E+k_R^2)=0$. Proposition~\ref{prop-specL}
reads as
\begin{thm} Let $\omega\in\dfrac{\pi}{2}+\pi\ZZ$ and $k_R\in\pi \ZZ$, then the spectrum
of $L$ consists of the real numbers $E$ satisfying at least one of the following conditions:
\begin{subequations}
\begin{gather}
         \label{eq-loc}
\Big(c(1;E+k_R^2)+\dfrac{\lambda}{6}\,s(1;E+k_R^2)\Big)\cdot\Big(s'(1;E+k_R^2)+\dfrac{\mu}{3}\, s(1;E+k_R^2)\Big)=\dfrac13,\\
         \label{eq-loc2}
s'(1;E+k_R^2)+\dfrac{\mu}{3}s(1;E+k_R^2)=0,\\
         \label{eq-loc3}
s(1;E+k_R^2)=0.
\end{gather}
\end{subequations}
Each point of the spectrum is an infinitely degenerate eigenvalue.
\end{thm} 
Note that the eigenvalues~\eqref{eq-loc} are the most interesting ones, as they arise
as the limit of the continuous spectrum. The Dirichlet eigenvalues~\eqref{eq-loc3}
are usually neglected in the physical works.

In the simplest case, when the scalar potential is zero
and the couplings are trivial, i.e. $U=0$, $\lambda=\mu=0$, one has
$s(x;E)=\dfrac{1}{\sqrt E}\,\sin\sqrt{E}x$, $c(x;E)=\cos \sqrt E x$,
and Eq.~\eqref{eq-loc} takes the form
\[
\cos^2 \sqrt{E+k_R^2}\in\big\{0,\dfrac{1}{3},1\big\},
\]
which was previously obtained in~\cite{vid1} for the case $k_R=0$.

We remark that the presence of the extreme localization is periodic with respect to
the shifts $k_R\mapsto k_R+\pi$, but not the energy levels themselves, as the functions
$s(\cdot;E+k_R^2)$ etc. are not periodic with respect to the Rashba constant.
It is worthwhile to note that the above results hold for any potential $U$
and any coupling constants $\lambda$ and $\mu$.

\subsection{Magneto-spin induced localization}

Another interesting situation appears at $\cos k_R=0$, i.e.
at $k_R\in \dfrac{\pi}{2}+\pi\ZZ$. In this case one has
\[
A^*A=6-2\begin{pmatrix}
\cos\big(\omega-\dfrac\pi 3\big)\,\Delta & 0\\
0 & \cos\big(\omega+\dfrac\pi 3\big)\,\Delta
\end{pmatrix}.
\]

For the values $\omega\in-\dfrac{\pi}{6}+\pi\ZZ$
the first component of $A^*A$ degenerates. In particular, any function of the form
$(f,0)$, $f\in \ell^2(\ZZ^2)$ becomes an eigenfunction of $A^*A$. For $\omega\in\dfrac{\pi}{6}+\pi\ZZ$
the same holds for the functions $(0,f)$. 

For further analysis we calculate the spectrum of $\Delta$.
\begin{lemma}\label{lem-tpi2}
For $\omega\in\pm\dfrac{\pi}{6}+\pi\ZZ$ the spectrum of $\Delta$ is absolutely continuous
and covers the segments $\big[-2\sqrt 3, -\sqrt 3\big]$ and $\big[\sqrt 3,2\sqrt 3\big]$.
\end{lemma}
\begin{proof}
Consider the unitary transformation
\[
U:\ell^2(\ZZ^2)\ni \big(f(m,n)\big)\mapsto \big(\be^{3\bi\omega mn}f(m,n)\big)\in\ell^2(\ZZ^2).
\]
Clearly, $U$ is unitary, and the operator $\Hat\Delta:=U^*\Delta U$ has the form
\begin{align*}
\Hat\Delta f(m,n)=&\be^{-6\bi\omega m}f(m,n+1)+\be^{6\bi\omega m}f(m,n-1)+ f(m+1,n)+f(m-1,n)\\
&+\be^{-3\bi\omega(2m-1)}f(m-1,n+1)+\be^{3\bi\omega(2m+1)}f(m+1,n-1).
\end{align*}
The operator obtained has the same spectrum as $\Delta$, but is periodic with respect to the shifts
$n\mapsto n+1$ and can be studied using the Bloch analysis.
Making the Bloch substitution $f(m,n)=\be^{\bi n q}\xi_m$, where
$q\in[0,2\pi)$ is the quasimomentum, we observe that the spectrum of $\Hat\Delta$ is
the union of the spectra of operators $H(q)$ acting in $\ell^2(\ZZ)$ and defined by
\begin{multline*}
\big(H(q)\xi\big)_m=2\exp\Big[-3\bi \omega\big(m-\dfrac{1}{2}\big)+\bi \dfrac{q}{2}\Big]
\cos\Big[3\omega\big(m-\dfrac{1}{2}\big)-\dfrac{q}{2}\Big]\, \xi_{m-1}\\
+2\cos\big[6\omega m-q\big]\,\xi_m
+2\exp\Big[3\bi \omega\big(m+\dfrac{1}{2}\big)-\bi \dfrac{q}{2}\Big]
\cos\Big[3\omega\big(m+\dfrac{1}{2}\big)-\dfrac{q}{2}\Big]\, \xi_{m+1}.
\end{multline*}
The operators $H(q)$ are nothing but the Harper operators for the triangular lattice~\cite{CW}.
Note that for $\omega\in\pm\dfrac{\pi}{6}+\pi\ZZ$
all these operators are invariant under the shift $m\mapsto m+2$.
Therefore, substituting into the equation $H(q)\xi=E\xi$
a vector $\xi$ satisfying $\xi_m=\be^{\bi\theta}\xi_{m-2}$ for all $m$, where $\theta\in[0,2\pi)$ is another quasimomentum,
one arrives at a $2\times 2$ linear system for the components $\xi_0$
and $\xi_1$,
\begin{multline*}
\be^{\bi(3\omega/2+q/2)}\cos\big(\frac{3\omega}{2}+\dfrac q 2\big) \be^{-\bi\theta}\xi_1
+\cos q\, \xi_0\\
\shoveright{+\be^{\bi(3\omega/2-q/2)}\cos\big(\frac{3\omega}{2}-\dfrac q 2\big)\xi_1=\dfrac{E}{2}\xi_0,}\\
\shoveleft{\be^{-\bi(3\omega/2-q/2)}\cos\big(\frac{3\omega}{2}-\dfrac q 2\big) \xi_0
-\cos q\, \xi_1}\\+\be^{-\bi(3\omega/2+q/2)}\cos\big(\frac{3\omega}{2}+\dfrac q 2\big)\be^{\bi\theta}\xi_0=\dfrac{E}{2}\xi_1.
\end{multline*}
The condition for the determinants to vanish takes the form
\begin{align*}
\dfrac{E^2}{4}&=1+\cos^2q+\cos(q-\theta)\,\cos q\\
&=\dfrac34+\big(\cos q+\dfrac12\cos(q-\theta)\big)^2+\sin^2(q-\theta).
\end{align*}
Taking here all possible values of $q$ and $\theta$ we arrive at the conclusion.
\end{proof}

Lemma~\ref{lem-tpi2} means that for values $\omega$ in question,
the spectrum of $A^*A$ has a continuous part, which is the union of the segments $[0,3]$ and $[9,12]$,
and an infinitely degenerate eigenvalue $6$.

Therefore, we arrive, as in subsection~\ref{sec-loc-mag}, to a series of infinitely
degenerate eigenvalues $E$ satisfying the same equation~\eqref{eq-loc} (i.e. the \emph{same} eigenvalues
as in the extreme localization case), which are isolated in the spectrum,
but we have additionally bands of  continuous spectrum given by
\[
\Big(c(1;E+k_R^2)+\dfrac{\lambda}{6}\,s(1;E+k_R^2)\Big)\cdot\Big(s'(1;E+k_R^2)+\dfrac{\mu}{3}\, s(1;E+k_R^2)\Big)
\in [0,\dfrac16]\cup[\dfrac 12,\dfrac23].
\]
In particular, for the free case with zero coupling constants one has the following characterization for $E$ to be in the spectrum of $L$:
\[
\cos^2\sqrt{E+k_R^2}\in \big[0,\dfrac16\big]\cup\big\{\dfrac13\big\}\cup\big[\dfrac 12,\dfrac23\big]
\cup\big\{1\big\}.
\]
The localization effect described in this subsection seems to be not covered by the existing works, and it would be interesting to know whether it can be really observed.
As for different values of the magnetic parameteres we have completely different eigensubspaces of $A^*A$, we conjecture that this localization mechanism can be used to control the spin polarization by the magnetic field, but this needs a further analysis.

\section{Appendix. Supersymmetric spectral analysis}

Here we prove the following proposition.

\begin{prop}\label{prop-susy}
Let $\HH_1$, $\HH_2$ are some Hilbert spaces,
$A$ be a bounded linear operator
from $\HH_1$ to $\HH_2$, and $m\in\RR$. On $\HH_1\oplus \HH_2$ consider the operator
\[
L=\begin{pmatrix}m & A^*\\
A & -m
\end{pmatrix}.
\]
Then
\begin{equation}
        \label{eq-spL}
\spec L=-\sqrt{\spec(AA^*+m^2)}\cup \sqrt{\spec(A^*A+m^2)}.
\end{equation}
\end{prop}

This proposition is formulated (without proof) in \cite{ogu} and is nothing but
an abstract version of proposition~2.5 in \cite{shw}; we give here a complete proof just for the sake of completeness.

\begin{proof}
First note that $\spec AA^*\setminus\{0\}=\spec A^*A\setminus\{0\}$ \cite{deift}.
Clearly, 
\begin{equation}
      \label{eq-LAA}
L^2=\begin{pmatrix}
A^*A+m^2 & 0\\
0 & AA^*+m^2
\end{pmatrix}.
\end{equation}
Therefore, $\spec L^2\setminus\{\pm m\}=\spec (AA^*+m^2)\setminus\{m^2\}$,
and for any $\lambda\in \spec AA^*\setminus\{0\}\equiv \spec AA^*\setminus\{0\}$ 
at least one of the numbers $-\sqrt{\lambda+m^2}$, $\sqrt{\lambda+m^2}$ lies in $\spec L$.
Let us show that actually they both are in the spectrum of $L$.

Let $\lambda>0$, $\lambda\in\spec A^*A$, then there exist a sequence $(\phi_n)$,
$\phi_n\in\HH_1$ such that
$\|\phi_n\|\ge 1$ and $\lim(A^*A-\lambda)\phi_n=0$. Denote
\begin{equation}
\psi_n:=\bigg[\lambda+\big(\sqrt{\lambda+m^2}-m\big)\begin{pmatrix}0 & A^*\\A & 0\end{pmatrix}\bigg]
\begin{pmatrix}
\phi_n\\0
\end{pmatrix}
\end{equation}
Clearly,
\[
\begin{pmatrix}
\phi_n\\0
\end{pmatrix}
\perp
\begin{pmatrix}
0 & A^*\\A & 0\end{pmatrix}
\begin{pmatrix}
\phi_n\\0
\end{pmatrix},
\]
which implies
\begin{equation}
     \label{eq-ll}
\|\psi_n\|\ge \lambda\|\phi_n\|\ge\lambda.
\end{equation}
By direct calculation,
\[
(L-\sqrt{\lambda+m^2}\,)\psi_n=\big(\sqrt{\lambda+m^2}-m\big)
\begin{pmatrix}
(A^*A-\lambda)\phi_n\\0
\end{pmatrix}.
\]
Therefore, $\lim (L-\sqrt{\lambda+m^2}\,)\psi_n= 0$. Together with \eqref{eq-ll} this implies
$\sqrt{\lambda+m^2}\in\spec L$.

To show $-\sqrt{\lambda+m^2}\in\spec L$ one has to consider the functions
\[
\psi_n:=\bigg[\lambda-\big(\sqrt{\lambda+m^2}-m\big)\begin{pmatrix}0 & A^*\\A & 0\end{pmatrix}\bigg]
\begin{pmatrix}
0\\ \phi_n
\end{pmatrix},
\]
where $\|\phi_n\|\ge 1$ and $\lim(AA^*-\lambda)\phi_n=0$ and to repeat the above steps.
To finish the proof of Eq.~\eqref{eq-spL} it is necessary to study the points $\pm m$.

For $m=0$, Eq.~\eqref{eq-LAA}
reads as $\spec L^2=\spec AA^*\cup\spec A^*A$,
and the conditions $0\in\spec L$
and $0\in\spec AA^*\cup\spec A^*A$ are equivalent.

Assume $m\ne 0$ and $m\in\spec L$, then there exist sequences
$(\phi_n)\in\HH_1$, $(\varphi_n)\in\HH_2$ with
\begin{equation}
        \label{eq-pp}
 \|\phi_n\|+\|\varphi_n\|\ge 1
\end{equation}
 and
\begin{equation}
    \label{eq-loc1}
\lim (L-m)\begin{pmatrix}
\phi_n\\
\varphi_n
          \end{pmatrix}\equiv
\lim\begin{pmatrix}
 A^*\varphi_n\\
A\phi_n -2m \varphi_n
\end{pmatrix}=0.
\end{equation}
Clearly, this implies $\lim A^*A\phi_n=0$.
Assume that $\lim\phi_n=0$, then \eqref{eq-loc1} shows $\lim \varphi_n=0$ which contradicts~\eqref{eq-pp}.
Therefore, there exists a subsequence $(\phi'_n)$ of $(\phi_n)$ such that
$\|\phi'_n\|\ge\epsilon$ for some $\epsilon>0$. Together with $\lim A^*A\phi'_n=0$
this implies $0\in\spec A^*A$.

Assume now $0\in\spec A^*A$, then there is a sequence $(\phi_n)\in \HH_1$ with 
$\|\phi_n\|\ge 1$ and $\lim \langle A^*A\phi_n,\phi_n\rangle\equiv \lim \|A\phi_n\|=0$. Then
\[
\lim(L-m)\begin{pmatrix}\phi_n\\0
\end{pmatrix}=\lim
\begin{pmatrix}
0 \\
A\phi_n
\end{pmatrix}=0,
\]
from which $m\in\spec L$.

The relationship between the conditions $-m\in L$
and $0\in\spec AA^*$  can be proved in a completely similar way.
\end{proof}

It may be useful to have an alternative formulation of proposition~\ref{prop-susy}.
\begin{corol}\label{corol-susy}
There holds
\begin{equation*}
\begin{aligned}
\spec L\setminus\{-m,m\}&=-\sqrt{\spec(AA^*+m^2)}\cup
\sqrt{\spec(AA^*+m^2)}\setminus\{-m,m\}\\
&\equiv -\sqrt{\spec(A^*A+m^2)}\cup
\sqrt{\spec(A^*A+m^2)}\setminus\{-m,m\}.
\end{aligned}
\end{equation*}
Furthermore, for $m\ne 0$ one has: $m\in\spec L$ iff $0\in\spec A^*A$,
$-m\in\spec L$ iff $0\in\spec AA^*$,
and for $m=0$ there holds
$0\in\spec L$ iff $0\in\spec A^*A\cup\spec AA^*$.
\end{corol}

\section*{Acknowledgments}
Numerous discussions with Vladimir Geyler
and Denis Bulaev are gratefully acknowledged.

The research was supported by the Deutsche Forschungsgemeinschaft
(fellowship PA 1555/1-1)
and the German--New Zealand cooperation of BMBF 
(project no. NZL 05/001 of the International
Bureau at the German Aerospace Center)
and ISAT (grant no. 9144/360402 of the Royal Society of New Zealand).
A large part of the work was done during the stay
at the Department of Mathematics
of the University of Auckland in March 2006. The author thanks
Boris Pavlov for valuable interaction and the warm hospitality
during the visit. 


\

\begin{thebibliography}{99}


\bibitem{p1} Abilio,~C.~C., Butaud,~P., Fournier,~Th.,
Pannetier,~B., Vidal,~J., Tedesco,~S., Dalzotto,~B.: \emph{Magnetic field
induced localization in a two-dimensional superconducting wire
network,} Phys. Rev. Lett. {\bf 83} (1999) 5102--5105.


\bibitem{ae}
Aeppli,~G., Chandra,~P.: \emph{Seeking a simple complex system,}
Science {\bf 275} (1997) 177--178.

\bibitem{akk} Akkermans,~E., Comtet,~A., Desbois,~J., Montambaux,~G., Texier,~C.:
\emph{Spectral determinant on quantum graphs,} Ann. Phys. (New York) {\bf 284} (2000) 10--51.

\bibitem{alx} Alexander,~S.: \emph{Superconductivity of networks.
A percolation approach to the effects of disorder,} Phys. Rev. B {\bf 27} (1983)
1541--1557.



\bibitem{jvb} von Below,~J.: \emph{A characteristic equation associated to an eigenvalue problem on $c\,^2$-networks,} Linear Algebra Appl. {\bf 71} (1985) 309--325.

\bibitem{jvb2} von Below,~J.:
\emph{Sturm-Liouville eigenvalue problems on networks,} Math. Meth. Appl. Sci. {\bf 10} (1988) 383--395.


\bibitem{r1}
Bercioux,~D., Governale,~M., Cataudella,~V., Ramaglia,~V.~M.:
\emph{Rashba-effect-induced localization in quantum networks,}
Phys. Rev. Lett. {\bf 93} (2004) 056802.

\bibitem{r2}
Bercioux,~D., Governale,~M., Cataudella,~V., Ramaglia,~V.~M.:
\emph{Rashba effect in quantum networks,}
Phys. Rev. B {\bf 72} (2005) 075305.

\bibitem{BirSus}
Birman,~M.~Sh., Suslina,~T.~A.:
\emph{A periodic magnetic Hamiltonian with a variable metric. The problem of absolute continuity,} St. Petersburg Math. J. {\bf 11} (2000) 203--232.

\bibitem{bol2}
Bolte,~J., Harrison,~J.:
\emph{Spectral statistics for the Dirac operator on graphs,}
J. Phys. A: Math. Gen. {\bf 36} (2003) 2747--2769.


\bibitem{bol1} Bolte,~J., Harrison,~J.:
\emph{The spin contribution to the form factor of quantum graphs,}
J. Phys. A: Math. Gen. {\bf 36} (2003) L433--L440.

\bibitem{bgp} Br\"uning,~J., Geyler,~V., Pankrashkin,~K.:
\emph{Cantor and band spectra for periodic quantum graphs with magnetic fields,}
Commun. Math. Phys. (to appear), Preprint math-ph/0511057.

\bibitem{bul} Bulla,~W., Trenkler,~T.:
\emph{The free Dirac operator on compact and non-compact
graphs,} J. Math. Phys. {\bf 31} (1990) 1157--1163.


\bibitem{br2}
 Bychkov,~Yu.~A.,  Rashba,~E.~I.:
\emph{Properties of a 2D electron gas with lifted spectral degeneracy,}
Sov. Phys. JETP Lett. {\bf 39} (1984) 78--80.

\bibitem{castro}
Castro,~J.~I.,  L\'opez,~A.:
\emph{The de Gennes-Alexander theory of superconducting micronetworks,}
In: Berger,~J., Rubinstein,~J. (Eds.):
\emph{Connectivity and Superconductivity}
 (Lecture Notes Phys. Monographs, vol. 62,
Springer, Berlin, 2000)   23--62 .


\bibitem{CC} Cattaneo, C.: \emph{The spectrum of the continuous Laplacian
on a graph.} Monatsh. Math. {\bf 124} (1997) 215--235.

\bibitem{MC2} 
Chaplik,~A.~V., Magarill,~L.~I.,:
\emph{Bound states in a two-dimensional short range potential induced by the spin-orbit interaction,} Phys. Rev. Lett. {\bf 96} (2006) 126402.


\bibitem{CW} Claro,~F.~H., Wannier,~G.~H.:
\emph{Magnetic subband structure of electrons in hexagonal lattices,}
Phys. Rev. B {\bf 19} (1979) 6068--6074.

\bibitem{dk} 
Debald,~S., Kramer,~B.: \emph{Rashba effect and magnetic field in semiconductor quantum wires,}
Phys. Rev. B {\bf 71} (2005) 115322.


\bibitem{deift} Deift,~P.:
\emph{Applications of a commutation formula,}
Duke Math. J. {\bf 45} (1978) 267--310.

\bibitem{dm} Derkach,~V.~A., Malamud,~M.~M.:
 \emph{Generalized resolvents
and the boundary value problems for Hermitian operators with gaps,}
J.~Funct. Anal. \textbf{95} (1991) 1--95.

\bibitem{des} Desbois, J.: \emph{Spectral determinant on graphs with generalized boundary conditions,}
Eur. Phys. J. B {\bf 24}  (2001) 261--266.
 
\bibitem{exdual} Exner,~P.: \emph{A duality between Schr\"odinger operators on graphs and certain Jacobi matrices,} Ann. Inst. Henri Poincar\'e Phys. Th\'eor. {\bf 66} (1997) 359--371.


\bibitem{exner-lkp} Exner,~P.: \emph{Lattice Kronig-Penney models,}
Phys. Rev. Lett.  {\bf 74} (1995) 3503--3506.


\bibitem{dGe} de~Gennes,~P.-G.: \emph{Diamagn\'{e}tisme de grains supraconducteurs
pr\`{e}s d'un seuil de percolation,} C.~R. Acad. Sci. Paris S\'{e}r.~II
{\bf292} (1981) 9--12.


\bibitem{gm} Geyler,~V.~A., Margulis,~V.~A.:
\emph{Anderson localization in the nondiscrete Maryland model,}
    Theor. Math. Phys. {\bf70} (1987) 133--140.



\bibitem{gna} Gnutzmann,~S., Altland,~A.:
\emph{Spectral correlation of individual quantum graphs,}
Phys. Rev. E {\bf 72} (2005) 056215.

\bibitem{KS}  Kostrykin,~V., Schrader,~R.:
\emph{Kirchhoff's rule for quantum wires,} J. Phys. A: Math. Gen. \textbf{32} (1999) 595--630.

\bibitem{KSM} Kostrykin,~V., Schrader,~R.: \emph{Quantum wires with magnetic fluxes,}
Commun. Math. Phys. {\bf 237} (2003) 161--179.

\bibitem{KoSm} Kottos,~T., Smilansky,~U.:
\emph{Periodic orbit theory and spectral statistics for quantum graphs,}
Ann. Phys. (New York) {\bf 274} (1999) 76--124.

\bibitem{kuch} Kuchment,~P.:
\emph{Graph models for waves in thin structures,}
Waves Random Media {\bf 12} (2002) R1--R24.


\bibitem{Ku1} Kuchment,~P.: \emph{Quantum graphs} I. \emph{Some basic structures,}
Waves Random Media {\bf14} (2004) S107--S128.

\bibitem{Ku2} Kuchment,~P.: \emph{Quantum graphs} II. \emph{Some spectral properties of
quantum and combinatorial graphs,} J. Phys. A: Math. Gen. {\bf 38} (2005) 4887--4900.

\bibitem{MC} Magarill,~L.~I., Chaplik,~A.~V.:
\emph{Spin-dependent electron localization in crystals,}
JETP Lett. {\bf 81} (2005) 162--166.


\bibitem{mpp} Mikhailova,~A., Pavlov,~B., Prokhorov,~L.:
\emph{Modelling of quantum networks,} Preprint math-ph/0312038.


\bibitem{mo2}
 Mills,~R.~G.~J.,  Montroll,~E.~W.: \emph{Quantum theory on a network} II. \emph{A solvable model which may have several bound states per node point,} J. Math. Phys.
{\bf 11} (1970) 2525--2538.

\bibitem{mo1}  Montroll,~E.~W.: \emph{Quantum theory on a network} I. \emph{A solvable model whose eigenfunctions are elementary functions,} J. Math. Phys. {\bf 11} (1970)
635--648.

\bibitem{mc} Moulopoulos,~K., Constantinou,~M.: \emph{Magnetic-field-induced localization in networks
with the $T_3$ geometry,} Phys. Lett. A {\bf 302} (2002) 39--47.

\bibitem{nap} Naud,~C.:
\emph{Transport quantique dans des nanostructures,}
Ann. Phys. (Paris) {\bf 27} no.~5 (2002) 1--140.

\bibitem{naud1} Naud,~C., Faini,~G., Mailly,~D.: \emph{Aharonov-Bohm cages in 2D normal metal networks,}
Phys. Rev. Lett. {\bf 81} (1998) 5888--5891.

\bibitem{naud2} Naud,~C., Faini,~G., Mailly,~D., Vidal,~J., Dou\c{c}ot,~B., Montambaux,~G., Wieck,~A., Reuter,~D.: \emph{Aharonov-Bohm cages in the GaAlAs/GaAs system,} Physica E {\bf 12} (2002) 190--196.

\bibitem{nic} Nicaise,~S.: \emph{Spectre des r\'eseaux topologiques finis,}
Bull. Sci. Math. {\bf 111} (1987) 401--413.

\bibitem{ogu} Ogurisu,~O.: \emph{Supersymmetric analysis of the spectral theory
on infinite graphs,} Preprint 02-242 on \url{http://www.ma.utexas.edu/mp_arc/}

\bibitem{kp} Pankrashkin,~K.: \emph{Spectra of Schr\"odinger operators on equilateral quantum graphs,}
Lett. Math. Phys. {\bf 77} (2006) 139--154.

\bibitem{kp2} Pankrashkin,~K.: \emph{Spectral duality for discrete and continuous Schr\"odinger operators on graphs and hypergraphs,} in preparation.

\bibitem{pavlov} Pavlov,~B.~S.: \emph{The theory of extensions and
explicitly solvable models,} Russian Math. Surveys
\textbf{42} (1987) 127--168.

\bibitem{ra}
Rashba,~E.~I.:
\emph{Properties of semiconductors with an extremum loop.} 1.
\emph{Cyclotron and combinational resonance in a magnetic field
perpendicular to the plane of the loop,}
Sov. Phys. Solid State {\bf 2} (1960) 1109--1122. 

\bibitem{roth} Roth, J.-P.: \emph{Spectre du laplacien sur un graphe,}
C. R. Acad. Sci. Paris {\bf 296} (1983) 783--795.

\bibitem{shw} Shigekawa,~I.:
\emph{Spectral properties of Schr\"odinger operators
with magnetic fields for a spin $\frac12$ particle,}
J. Funct. Anal. {\bf 101} (1991) 255--285.

\bibitem{asob} Sobolev,~A.~V.:  \emph{Absolute continuity of the periodic magnetic Schr\"odinger operator,} Invent. Math. {\bf137} (1999) 85--112. 

\bibitem{suth} Sutherland,~B.: \emph{Localization of electronic wave functions due to local topology,}
Phys. Rev. B {\bf 34} (1986) 5208--5211.


\bibitem{vid2} Vidal,~J., Butaud,~P., Dou\c{c}ot,~B., Mosseri,~R.:
\emph{Disorder and interactions in Aharonov-Bohm cages,}
Phys. Rev. B {\bf 64} (2001) 155306.

\bibitem{deloc} Vidal,~J., Dou\c{c}ot,~B., Mosseri,~R., Butaud,~P.:
\emph{Interaction induced delocalization for two particles in a periodic potential,}
Phys. Rev. Lett. {\bf 85} (2000) 3906--3909.

\bibitem{vid1} Vidal,~J., Mosseri,~R., Dou\c{c}ot,~B.:
\emph{Aharonov-Bohm cages in two-dimen\-sional structures,} Phys. Rev. Lett. {\bf 86} (2001) 5104--5107.


\end{thebibliography}
\end{document}